\documentclass[a4paper]{jpconf}
\usepackage{graphicx}
\begin{document}
\title{Two-orbital view on the origin of the material dependence of $T_c$ in the single-layer cuprates}

\author{Hirofumi Sakakibara$^1$, Hidetomo Usui$^2$, Kazuhiko Kuroki$^1$, Ryotaro Arita$^3$, and Hideo Aoki$^4$}

\address{$^1$Department of Engineering Science,The University of Electro-Communication, Tokyo, Japan}
\address{$^2$Department of Applied Physics,The University of Electro-Communication, Tokyo, Japan}
\address{$^3$Department of Applied Physics,The University of Tokyo, Tokyo, Japan}
\address{$^4$Department of Physics,The University of Tokyo, Tokyo, Japan}

\ead{hiro\_rebirth@vivace.e-one.uec.ac.jp}
\begin{abstract}
Using the $d_{x^2-y^2}$+$d_{z^2}$ two orbital model of the high $T_c$ cuprates
 obtained from 
the first-principle calculation, we show 
that the material dependence of the Fermi surface shape 
can be understood by the 
degree of the mixture between the $d_{x^2-y^2}$ and the $d_{z^2}$ orbitals.
We explain, through investigating the tightbinding hopping integrals, 
why some cuprates have square shaped Fermi surface, while others 
have more rounded ones.
From this viewpoint, we explain the experimentally observed 
correlation between the curvature of the 
Fermi surface and $T_c$. 
\end{abstract}

\section{Introduction}

In the high $T_c$ cuprates, the 'main band' having strong 
Cu$3d_{x^2-y^2}$ orbital character constructs the Fermi surface, 
and the single band model that considers only the main band 
has often been adopted in the theoretical studies.
Such single band or related models have provided us 
many understandings, but there 
still remain unresolved problems.
One of the issues often discussed with controversy is the relationship 
between the curvature of the 
Fermi surface and the critical temperature ($T_c$). 
It is well known that even within the single-layered cuprates, there 
is significant difference of $T_c$, 
for example, the La compound with $T_c \simeq 40$K and the 
Hg compound with $T_c \simeq 90$K. 
From the band structure point of view, the La cuprate has 
relatively square shaped (diamond-like) Fermi surface, 
while the Hg material has a more round one. 
In fact, it has been recognized that low $T_c$ materials have 
square shaped Fermi surface, while 
high $T_c$ ones tend to have round Fermi surface\cite{Pavarini,Tanaka}.
Although some phenomenological\cite{Moriya} or 
$t-J$ model\cite{Shih,Prelovsek} studies describe 
such tendency, a number of Hubbard-type-model studies with 
realistic values of the on-site $U$  have not succeeded in 
reproducing such a tendency\cite{Scalapino}.
For example, the dynamical cluster 
approximation studies that adopt a single band model
\cite{Maier}, or a more realistic three band model\cite{Kent} that considers 
$p_{\sigma}$ orbitals  show the opposite tendency.

To give insight into this long standing problem, 
we have constructed a $d_{x^2-y^2}+d_{z^2}$ two orbital model 
that considers all the orbitals that have $e_g$ symmetry\cite{H.S.}.  
In this proceedings, we  focus on the relationship 
between the conventional single orbital model and the two-orbital
model, and explain the difference of the parameters that gives the 
material dependence.

\section{$d_{x^2-y^2}+d_{z^2}$ two-orbital model}

In Fig.1, we show 
the band dispersion of the two-orbital model constructed by exploiting 
maximally localized Wannier orbitals,\cite{MaxLoc} 
which are obtained from the first-principles calculation results\cite{pwscf}.
The thickness of the lines represents the strength of the respective 
orbital character. 
In the La compound, a significant amount 
of $d_{z^2}$ character is present in the 
main band around the wave vector 
$k=(\pi,0)(0,\pi)$(denoted as N in La and M  in Hg), while 
in the Hg compound such a mixture of the $d_{z^2}$ component 
is absent\cite{Shiraishi,Eto,Freeman,Andersen}.

\begin{figure}[h]
\includegraphics[width=16cm]{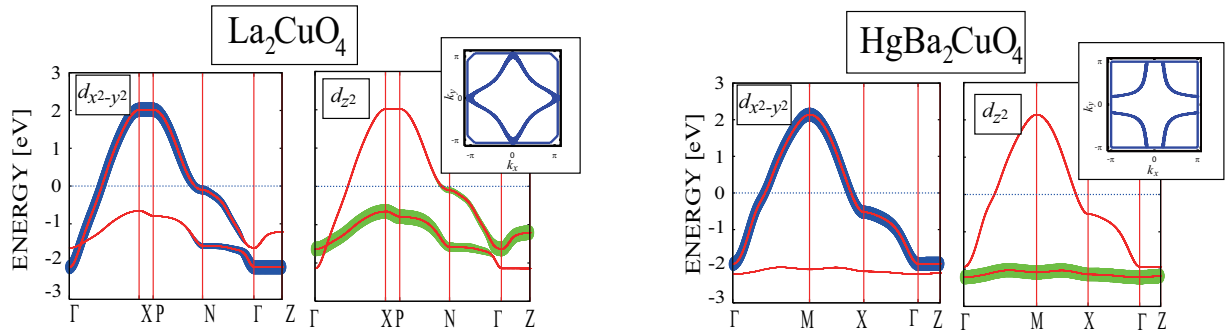}\hspace{2pc}%
\caption{\label{fig1}The band structure of the 
two orbital model for La$_2$CuO$_4$ (left) and HgBa$_2$CuO$_4$(right). 
The insets depict the Fermi surfaces (for the total band filling $n=2.85$). 
The thickness of the lines represents the strength of the 
respective orbital character.}
\end{figure}

\section{Comparison from the viewpoint of the hopping integrals}


Let us now look into the hopping integrals of the models.
In table I, the hopping integrals within the $d_{x^2-y^2}$ 
orbitals of the two orbital model are displayed for the La and Hg compounds.
For comparison, we also show the values of the 
single band model obtained by the same method.
Conventionally, the difference of the Fermi surface 
curvature is represented by the value
of the second nearest neighbor 
hopping $t_2$ and the third nearest hopping $t_3$, where 
large $|t_2|$ and $|t_3|$ gives more rounded Fermi surface.  
This tendency is in fact seen in the table.

\begin{table}[!h]
\caption{Hopping integrals 
within the $d_{x^2-y^2}$ orbital 
for the single and two-orbital models, 
and $\Delta E \equiv E_{x^2-y^2}-E_{z^2}$.}
\label{hop-para}
\begin{tabular}{ c| l l l l}
\hline
 & Single-orbital & & \hspace{1.0cm} Two-orbital & \\
 &  La & Hg  &\hspace{1.0cm}  La & Hg\\ \hline
$t_1 \rm{[eV]}$ & -0.444 & -0.453 & \hspace{1.0cm} -0.471 & -0.456 \\ 
$t_2 \rm{[eV]}$ & 0.0284 & 0.0874 & \hspace{1.0cm} 0.0932 & 0.0993 \\
$t_3 \rm{[eV]}$ & -0.0357 & -0.0825  & \hspace{1.0cm}  -0.0734 & -0.0897 \\
$(|t_2|+|t_3|)/|t_1|$ & 0.14 &  0.37 & \hspace{1.0cm} 0.35 & 0.41 \\ \hline 
$\Delta E\rm{[eV]}$ & -  &  - &\hspace{1.0cm}  0.91 & 2.19 \\
\hline
\end{tabular}
\end{table}


If we turn to the hopping integrals of the two orbital model, 
the values shown in Table I is surprising in that 
the $d_{x^2-y^2}$ orbital has 
large $|t_2|$ and $|t_3|$ even for the La compound, just as in the Hg compound.
Namely, the curvature of Fermi surface is not governed by the 
$d_{x^2-y^2}$ distant hoppings in the two orbital model that 
considers the $d_{z^2}$ orbital explicitly, 
and another parameter plays an important role.
In fact, the parameter is the on-site energy difference 
$\Delta E\equiv E_{x^2-y^2}-E_{z^2}$.
In other words, $\Delta E$ in the two-orbital model 
determines the $(|t_2|+|t_3|)/|t_1|$ ratio in the effective 
single-orbital model. This can be understood as follows.

\begin{figure}[h]
\includegraphics[width=16cm]{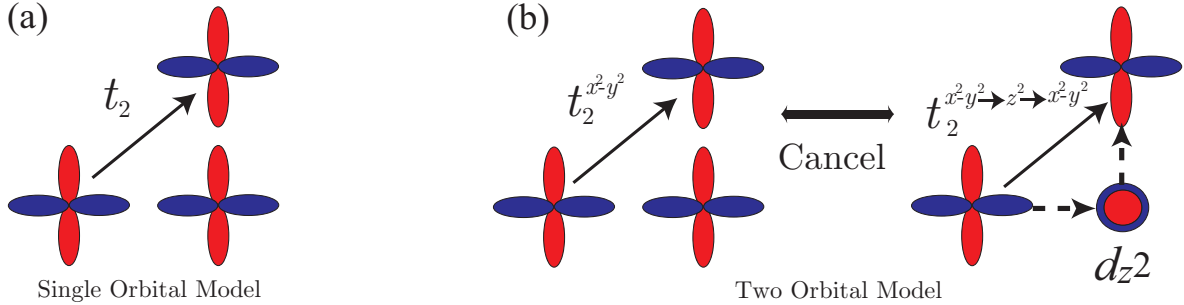}\hspace{2pc}%
\caption{\label{fig2}A  schematic image of the $d_{x^2-y^2}$
-$d_{x^2-y^2}$ diagonal 
hopping within the single (a) and two-orbital(b) models. 
The left side of (b) shows 
the direct path, while the right side is the indirect path 
via the $d_{z^2}$ orbital.}
\end{figure}

An electron can hop from a $d_{x^2-y^2}$ orbital 
to its second nearest neighbor 
mainly via the direct diagonal $d_{x^2-y^2}$-$d_{x^2-y^2}$ 
path or the indirect 
$d_{x^2-y^2} \rightarrow d_{z^2} \rightarrow d_{x^2-y^2}$ path.
The latter can be considered as a second order perturbation process 
when the $d_{z^2}$ orbital degrees of freedom is integrated out to obtain 
the effective single band model.
As shown in table I, the value of the direct hopping is nearly equal 
between the La and the Hg compound, 
so that the material dependence comes from the indirect term.


The amplitude of the indirect term is generally large 
for smaller $\Delta E$.
In the La compound, the relatively small $\Delta E$ mainly enhances the 
indirect path $d_{x^2-y^2} \rightarrow d_{z^2} \rightarrow d_{x^2-y^2}$.
From the first-principles result, the direct hopping and the indirect term 
are found to have the opposite sign, so 
the cancellation is strong when the amplitude of the indirect path is strong.
As a result, the small $\Delta E$ for the La compound 
gives the small effective $t_2$ and $t_3$.


The above analysis shows that the curvature of the Fermi surface in  the 
La cuprate is 
suppressed due to the large mixture between $d_{x^2-y^2}$ and $d_{z^2}$ 
orbitals. Therefore, the strong $d_{z^2}$ mixture in the 
Fermi surface around the wave vectors $(\pi,0)/(0,\pi)$ 
is the origin of the weak curvature.
%

\section{The relationship between Fermi surface and $T_c$}

\begin{figure}[h]
\includegraphics[width=6cm]{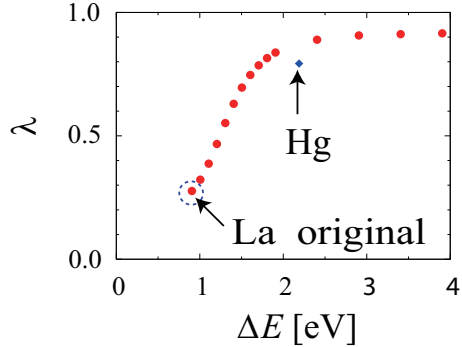}\hspace{2pc}%
\begin{minipage}[b]{14pc}
\caption{\label{fig3}The eigenvalue $\lambda$ of the 
Eliashberg equation for $d$-wave superconductivity plotted 
against $\Delta E\equiv E_{x^2-y^2}-E_{z^2}$. The points with 
arrows are the results obtained using the original lattice 
structure (determined experimentally). Circles are 
the results obtained by hypothetically varying $\Delta E$ in the La system.}
\end{minipage}
\end{figure}

Now, we finally discuss how $\Delta E$ affects the many body properties, 
especially $d$-wave superconductivity. We consider a many body 
Hamiltonian that considers the on-site multiorbital interactions with 
realistic values.
In Fig.3, we plot the $d$-wave 
eigenvalue of the linearized Eliashberg equation, 
where the Green's functions are obtained within the  
fluctuation exchange approximation\cite{Bickers,Dahm}. 
The eigenvalue reaches unity when $T=T_c$, so  $\lambda$
at a fixed temperature (here $T=0.01$[eV]) 
can be used as a qualitative measure for $T_c$.
In the figure, the points indicated by the arrows are 
the results of the model constructed from the experimentally 
determined lattice parameters\cite{La-st,Hg-st}.
This shows that the Hg compound has higher $T_c$ than the 
La compounds, so the result is consistent with the experiments. 
We also plot $\lambda$ when $\Delta E$ is varied hypothetically in the 
La system.
Note that $t_2^{d_{x^2-y^2}}$ and $t_3^{d_{x^2-y^2}}$ are 
fixed because in the first principles calculation they are not so much 
material dependent, as mentioned above.
From Fig.\ref{fig3}, it can be seen that 
$\lambda$ increases monotonically with $\Delta E$, 
and comes close to the value of 
the Hg result. 
From this result, we can say that $\Delta E$ governs both the 
shape of the Fermi surface (via the effective $t_2$ and $t_3$) 
and $T_c$.

\section{Conclusion}

In conclusion, we have shown that 
the degree of the orbital mixture controls the shape of the 
Fermi surface, and this gives the correlation between the Fermi surface 
shape and $T_c$. In this picture, the parameter that largely contributes to the 
material dependence is $\Delta E$, the on-site level offset 
between the $d_{x^2-y^2}$ and $d_{z^2}$ Wannier orbitals.
This picture also explains the conventionally adopted material 
dependence of the second and third nearest neighbor hoppings in the 
single band model. 

\section*{References}
\begin{thebibliography}{99}
\bibitem{Pavarini}
E. Pavarini {\it et al.}, Phys. Rev. Lett. {\bf 87}, 047003 (2001).
\bibitem{Tanaka} K. Tanaka {\it et al.}, Phys. Rev. B {\bf 70}, 092503 (2004).
\bibitem{Moriya} T. Moriya and K. Ueda, J. Phys. Soc. Jpn. {\bf 63}, 1871 
(1994).
\bibitem{Shih} 
C.T. Shih {\it et al.}, Phys. Rev. Lett. {\bf 92}, 227002 (2004). 
\bibitem{Prelovsek} P. Prelov$\rm \check{s}$ek and A. Ram$\rm \check{s}$ak, 
Phys. Rev. B {\bf 72}, 012510 (2005).
\bibitem{Scalapino} For a review, see D.J. Scalapino, 
{\it Handbook of High Temperature Superconductivity}, Chapter 13, 
Eds. J.R. Schrieffer and J.S. Brooks (Springer, New York, 2007).
\bibitem{Maier} Th. Maier {\it et al.}, Phys. Rev. Lett. {\bf 85}, 1524 (2000).
\bibitem{Kent} P.R.C. Kent {\it et al.}, Phys. Rev. B {\bf 78}, 035132 (2008).
\bibitem{H.S.} H. Sakakibara, H. Usui, K. Kuroki, R. Arita, and 
H. Aoki, Phys. Rev. Lett. {\bf 105}, 057003(2010) 
\bibitem{MaxLoc} N. Marzari and D. Vanderbilt, Phys. Rev. B 
{\bf 56}, 12847 (1997); 
I. Souza, N. Marzari and D. Vanderbilt, 
Phys. Rev. B {\bf 65}, 035109 (2001).
The Wannier functions are generated by the code developed by
A. A. Mostofi {\it et al.}, 
(http://www.wannier.org/).
\bibitem{pwscf} 
S. Baroni {\it et al.}, 
http://www.pwscf.org/.
Here we take the exchange correlation functional introduced by
J. P. Perdew {\it et al.}
[Phys. Rev. B {\bf 54}, 16533 (1996)], and the wave functions are expanded by 
plane waves up to a cutoff energy of 60 Ry with 
20$^3$ $k$-point meshes.
\bibitem{Shiraishi} K. Shiraishi {\it et al.}, Solid State Commun. {\bf 66},
629 (1988).
\bibitem{Eto} H. Kamimura and M. Eto, J. Phys. Soc. Jpn. {\bf 59}, 3053 
(1990); M. Eto and H. Kamimura, J. Phys. Soc. Jpn. {\bf 60}, 2311 (1991).
\bibitem{Freeman} A.J. Freeman and J. Yu, Physica B {\bf 150}, 50 (1988).
\bibitem{Andersen} O.K. Andersen {\it et al.}, J. Phys. Chem. Solids 
{\bf 56}, 1573 (1995).
\bibitem{Bickers} N.E. Bickers {\it et al.}, 
Phys. Rev. Lett. {\bf 62}, 961 (1989).
\bibitem{Dahm} T. Dahm and L. Tewordt, Phys. Rev. Lett. {\bf 74}, 793 (1995)
\bibitem{La-st}
J.D. Jorgensen {\it et al.}, Phys. Rev. Lett. {\bf 58}, 1024 (1987).
\bibitem{Hg-st}
J.L. Wagner {\it et al.}, Physica C {\bf 210}, 447 (1993).

\end{thebibliography}
\end{document}